\documentclass[aps,floatfix,showpacs,twocolumn]{revtex4}
\usepackage{amsmath,amssymb}
\usepackage{graphicx}
\usepackage{epsfig}
\usepackage{psfrag}
\usepackage{hyperref}

\newcommand{\bm}{\boldsymbol}

\begin{document}

\hsize\textwidth\columnwidth\hsize\csname@twocolumnfalse\endcsname

\title{Coherent Magneto-Optical Effects in Topological Insulators: \\
  Excitation Near the Absorption Edge}

\author{Wang-Kong Tse}
\affiliation{Department of Physics and Astronomy, Center for
Materials for Information Technology, The University of Alabama, Alabama 35487, USA}

\begin{abstract}
We study coherent optics in topological 
insulator surface states with broken time-reversal symmetry and
develop a theory for the dynamical Hall effect driven by intense
electromagnetic field. The influence of optical Stark effect 
enters as nonlinear dependence on the optical field in the resulting 
Faraday $\theta_F$ and Kerr $\theta_K$ rotations. This nonlinear
correction is found to decrease $\theta_F$ with
the strength of the \textit{A.C.} electric field, whereas $\theta_K$
exhibits a non-monotonic behavior. 
We also assess the 
effects of relaxation and dephasing on the Hall and
magneto-optical responses when the frequency detuning is comparable to
the inverse lifetime of the conduction electrons. 
\end{abstract}
\pacs{78.20.Ls,73.43.-f,75.85.+t,78.67.-n}

\maketitle


\section{Introduction} \label{sec:intro}
Topological insulators are materials with strong spin-orbit coupling
that host topologically protected and gapless surface
states \cite{TIPredict}. The surface states' electrons are massless Dirac fermions with 
a Dirac cone energy dispersion close to the band crossing point that
corresponds to spin degeneracy. One intriguing property that follows
from the removal of this spin degeneracy by time-reversal 
symmetry breaking is that a band gap opens up and the system exhibits a Hall 
response \cite{TIReview}. If the Fermi level falls within the band
gap, the Hall effect is quantized with a Hall conductivity equal to
half of the unit conductance $\sigma_{xy} = e^2/2h$. 
The half quantization is a consequence of the helical spin texture of the
surface Dirac electrons wrapping half of the unit sphere when the
lower surface band is fully filled.  
It may also be understood as a bulk magnetoelectric
effect \cite{QiPRB,Vanderbilt,SCZhangMO} developed from an additional term $\propto \theta \bm{E}\cdot\bm{B}$ in the
electromagnetic Lagrangian (where $\theta$ is called the axion coupling constant \cite{Wilczek}). 
As a result of this quantized Hall effect, topological insulator
surface states behave as a quantum anomalous Hall insulator (also
called a Chern insulator) and give rise to strong 
magneto-optical responses. The Faraday $\theta_F$ and Kerr $\theta_K$ angles are predicted to
display quantized values with $\theta_F$ quantized to multiples to
the fine structure constant $\alpha = 1/137$ and $\theta_K$ quantized
to a full-quarter rotation $\pi/2$ in topological insulator thin films
\cite{TI_MO1,TI_MO2,TI_MO3}. Soon after theoretical 
predictions, colossal values of up to 
$60^{\circ}$ of the Kerr 
effect \cite{Rolando} in Bi$_2$Se$_3$ topological insulator thin films
were measured. Recently, three groups have independently reported
measurements of the predicted quantization of the 
Faraday effect \cite{Fara1,Fara2,Fara3,Fara4}. 
The anomalous Hall transport is, naturally, a linear response effect
and therefore the resulting topological Faraday and Kerr angles are
independent of the optical field strength, corresponding to a regime
where the electromagnetic field can be treated as a perturbation. 

Strong \textit{A.C.} fields lead to interesting nonlinear optical
properties and are 
studied extensively in conventional semiconductors \cite{NLO}. 
For example, strong electromagnetic radiation acts to renormalize the
conduction and valence bands in the saturation state of a semiconductor
with the optically dressed electrons and
holes behaving as new quasiparticles \cite{Galitskii}. 
Optical nonlinearity and optically-induced coherent
effects pose a new frontier in recently discovered materials including
topological insulators, where coherent control of topological
properties might be possible. In particular, the influence of strong
radiation on the magneto-transport properties in topological
insulators presents an interesting and unexplored area of investigation 
that could lead to new insights on the interplay between light and band
topology. 

The present paper attempts to address some of the questions along this 
direction by generalizing our previous considerations on the
magneto-transport and magneto-optical properties of topological
insulators to strong electromagnetic fields. 
We shall focus on the small detuning regime where the light frequency is close to the
absorption threshold, so that the rotating wave approximation (RWA)
remains a viable strategy of solution while 
allowing the effects of strong
electromagnetic fields to be studied non-perturbatively. 

An outline of the paper is as follows. We first lay out our model for 
topological insulator surface states coupled to electromagnetic fields in Sec.~\ref{sec:model}. In
Sec.~\ref{sec:dressed}, we consider the coherent regime in the absence
of damping and discuss the effects of strong fields on the resulting non-equilibrium
quasiparticle distribution functions and energy dispersions. We then
proceed to calculate the dynamical longitudinal and Hall current 
responses in Sec.~\ref{sec:current}. In Sec.~\ref{sec:FaraKerr}, 
we formulate the equations that incorporates nonlinear optical
effects in the transmission coefficients and calculate the nonlinear magneto-optical
Faraday and Kerr effects. Finally, we consider the effects of relaxation and dephasing on the dynamical current responses
and magneto-optical effects in the regime where the detuning is
comparable to damping rates in Sec.~\ref{sec:relax}. 
  
\section{Model} \label{sec:model}
Under time-reversal symmetry breaking, topological insulator surface states are described by the $2\times2$ massive Dirac Hamiltonian 
\begin{equation}
H^0 =\frac{1}{2}\left(\varepsilon_{k+}-\varepsilon_{k-}\right)\begin{bmatrix}
\cos\theta_{k} &  e^{-i\phi_k}\sin\theta_{k} \\
e^{i\phi_k}\sin\theta_{k}  &
-\cos\theta_{k}\end{bmatrix}, \label{H0}
\end{equation}
where $\varepsilon_{k\pm} = \pm \alpha_k$  are the conduction (+) and valence band (-)
energies with $\alpha_k = \sqrt{(v  k)^2+{\Delta}^2}$, $\tan\phi_k = k_y/k_x$ is the azithmuthal angle of the electron
momentum, $\theta_{k}$ is the polar angle with $\cos\theta_{k} =
{\Delta}/{\alpha_{k}}$ and $\sin\theta_{k} = {v k}/{
  \alpha_{k}}$. Throughout this work, we set $\hbar = 1$ unless
otherwise specified.  
The discrete binary degrees of freedom describe
electron spins. $\Delta$ is the Zeeman field acting on the electron 
spins, which gives the Dirac surface states a band gap of
$2\Delta$. This Zeeman field can be induced by exchange coupling
to the topological insulator surface states or by doping with magnetic
atoms. 

We introduce a set of spin unit vectors 
$\{\hat{\boldsymbol{\alpha}}_{k},
\hat{\boldsymbol{\beta}}_{k},
\hat{\boldsymbol{\gamma}}_{k}\}$ (the reader is referred to the
Appendix for their definitions), and define the Pauli matrices in 
the new basis $(\sigma_{\alpha},\sigma_{\beta},\sigma_{\gamma}) = \bm{\sigma}\cdot(\hat{\boldsymbol{\alpha}}_{k},
\hat{\boldsymbol{\beta}}_{k},
\hat{\boldsymbol{\gamma}}_{k})$ that satisfy the commutation relation $\left[\sigma_{\alpha},\sigma_{\beta}\right] =
2i\sigma_{\gamma}$. $\hat{\boldsymbol{\alpha}}_{k}$ describes the
local orientation of the electron spin at momentum $\bm{k}$. The
Hamiltonian can then be expressed in the form of a Zeeman coupling
term $H^0 = \bm{\mathcal{B}}_{k}^0\cdot\bm{\sigma}/2$ with
$\bm{\mathcal{B}}_{k}^0 = 2\alpha_{k}\hat{\boldsymbol{\alpha}}_{k}$ taking
on the meaning of an effective magnetic field.  

Initially, our system is in an unexcited state with
a fully filled valance band and an empty conduction band. A linearly
polarized light is then illuminated onto the topological insulator 
with an electric field $\boldsymbol{E} = E_0 \cos\omega t
\hat{\boldsymbol{e}}_x$ and a polarization state indicated by the  unit
vector along $x$ direction $\hat{\boldsymbol{e}}_x$. The electron-photon
interaction Hamiltonian $H^{{p}}(t) = ({eE_0
  v}/{\omega})  \sin\omega t\sigma_x$ can be similarly expressed as a
Zeeman coupling term $H^{{p}}(t) =
\bm{\mathcal{B}}_{{k}}^{{p}}(t)\cdot\bm{\sigma}/2$ with a light-induced effective
magnetic field 
%
\begin{eqnarray}
\bm{\mathcal{B}}_{{k}}^{p} &=& \Lambda\sin\omega t ( 
  \sin\theta_k\cos\phi_k\hat{\boldsymbol{\alpha}}_{k}-
  \sin\phi_k\hat{\boldsymbol{\beta}}_{k} \nonumber \\
&&-\cos\theta_k\cos\phi_k\hat{\boldsymbol{\gamma}}_{k}),
\end{eqnarray}
where $\Lambda = {2eE_0 v}/{\omega}$ 
is the strength of the interband
transition matrix element and 
corresponds to the energy acquired by an electron driven by the
\textit{A.C.} field over a half period. 
%
%

We consider optical response of the system under irradiation with
intense off-resonant light having a frequency $\omega
< 2\Delta$ and a small detuning  $\delta \equiv
2\Delta-\omega \ll 2\Delta$. 
The regime of strong electromagnetic fields is characterized by $\Lambda \gg 1/\tau$, 
where $\tau$ is the electron lifetime. 
The dynamics of the system's $2\times 2$ density matrix 
$\rho_k(t)$ is governed by the quantum kinetic equation 
%
\begin{equation}
\frac{\partial
  \rho_k}{\partial t}+i\left[H,\rho_k\right] = I_{\mathrm{c}}(\rho_k,t), \label{QKE}
\end{equation}
where $H(t) = H^0+H^{{p}}(t)$ is the total Hamiltonian of the
system, and the collision integral 
$I_{\mathrm{c}}(\rho_k,t)$ takes into account damping effects from relaxation and
dephasing. In the following, (1) we first study and elucidate the main
physics in the coherent regime 
where the detuning $\delta \gg 1/\tau$ in Secs.~\ref{sec:dressed}-\ref{sec:FaraKerr}; (2) we then consider the case
$\delta \lesssim 1/\tau$ where the effects of relaxation and dephasing 
become non-negligible in  Sec.~\ref{sec:relax}. 

\section{Coherent Light-Driven Spin Dynamics} \label{sec:dressed}
In the coherent regime when detuning $\delta \gg 1/\tau$, one can ignore the collision
integral in the kinetic equation Eq.~(\ref{QKE}). To analyze the spin 
dynamics, we resolve the density matrix into its charge $n_k$ and spin 
$\bm{S}_k = {S}_k^{\alpha}\hat{\boldsymbol{\alpha}}_{k}+{S}_k^{\beta}\hat{\boldsymbol{\beta}}_{k}+{S}_k^{\gamma}\hat{\boldsymbol{\gamma}}_{k}$ sectors so that $\rho_k = n_k
\mathbb{I}+\bm{S}_k\cdot\bm{\sigma}/2$, where 
${S}_k^{\alpha},{S}_k^{\beta},{S}_k^{\gamma}$ are real. This gives the
Bloch equation governing the dynamics of the spin density matrix due
to the total effective magnetic field $\bm{\mathcal{B}}_{{k}} = 
\bm{\mathcal{B}}_{{k}}^0+\bm{\mathcal{B}}_{{k}}^{p}$,
%
\begin{equation}
\frac{\partial  \bm{S}_k}{\partial t}+\bm{S}_k\times \bm{\mathcal{B}}_{{k}} = 0,
\label{SKE}
\end{equation}
with the initial condition $\bm{S}_k(t=0) = -\hat{\boldsymbol{\alpha}}_{k}$. 
Eq.~(\ref{SKE}) implies that the magnitude of the $\bm{S}_k$
is a constant of motion with $\vert \bm{S}_k \vert = 1$. 
The component of the density matrix ${S}_k^{\alpha}$ along $\hat{\bm{\alpha}}_{k}$ corresponds to
population difference due to interband transitions whereas the
components ${S}_k^{\beta}, {S}_k^{\gamma}$ along the orthogonal
directions $\hat{\bm{\beta}}_{k}, \hat{\bm{\gamma}}_{k}$ corresponds to
interband coherence. 

It is convenient to define the longitudinal $\hat{\bm{\alpha}}_{k}^{\mathrm{L}} = \hat{\bm{\alpha}}_{k}$
and transverse $\hat{\bm{\alpha}}_{k}^{\mathrm{T}} = \hat{\boldsymbol{\beta}}_{k}-i
  \hat{\boldsymbol{\gamma}}_{k}$ spin vectors and transform our
  reference frame into the rotating frame at the laser frequency. 
In the rotating frame then, the longitudinal 
$S_k^{\mathrm{L}}$ and transverse spin density matrices
$S_k^{\mathrm{T}}$ are given as   
%
\begin{eqnarray}
S_k^{\mathrm{L}} &\equiv& \bm{S}_{k}\cdot
\hat{\bm{\alpha}}_{k}^{\mathrm{L}}, \label{SLdef} \\
S_k^{\mathrm{T}}
&\equiv& e^{i\omega t}\bm{S}_{k}\cdot
\hat{\bm{\alpha}}_{k}^{\mathrm{T}}, \label{STdef}
\end{eqnarray}
whereas the longitudinal $\mathcal{B}_{k}^{\mathrm{L}}$ and transverse
$\mathcal{B}_{k}^{\mathrm{T}}$ components of the total effective magnetic
field are defined as 
\begin{eqnarray}
\mathcal{B}_{k}^{\mathrm{L}} &=& 
\bm{\mathcal{B}}_k\cdot \hat{\bm{\alpha}}_{k}^{\mathrm{L}}-\omega, \label{BLdef} \\
\mathcal{B}_{k}^{\mathrm{T}} &=&
e^{i\omega t}\bm{\mathcal{B}}_k\cdot \hat{\bm{\alpha}}_{k}^{\mathrm{T}}. \label{BTdef}
\end{eqnarray}
It then follows from the above Eqs.~(\ref{BLdef})-(\ref{BTdef}) that 
%
\begin{eqnarray}
\mathcal{B}_k^{\mathrm{L}} &=& 2\alpha_k-\omega+\hat{\bm{\alpha}}_{k}^{\mathrm{L}}\cdot\bm{\mathcal{B}}_k^{p}, \label{BkL1} \\
\mathcal{B}_k^{\mathrm{T}} &=& e^{i\omega
      t}\hat{\bm{\alpha}}_{k}^{\mathrm{T}}\cdot\bm{\mathcal{B}}_k^{p}. \label{BkT1}
\end{eqnarray}
%
%
In terms of these new variables, we can recast
Eq.~(\ref{SKE}) as
\begin{eqnarray}
i\frac{\partial  {S}_k ^{\mathrm{L}}}{\partial t} &=&
\frac{1}{2}\left[\left({S}_k
    ^{\mathrm{T}}\right)^*\mathcal{B}_{k}^{\mathrm{T}}-{S}_k
    ^{\mathrm{T}}\left(\mathcal{B}_{k}^{\mathrm{T}}\right)^*\right], \label{SKE2}
\\
i\frac{\partial  {S}_k ^{\mathrm{T}}}{\partial t} &=&
{S}_k^{\mathrm{T}}\mathcal{B}_{k}^{\mathrm{L}}-{S}_k^{\mathrm{L}}\mathcal{B}_{k}^{\mathrm{T}}, \label{SKE3}
\end{eqnarray}
supplemented by the initial conditions $ {S}_k ^{\mathrm{L}}(t= 0) = -1$
and  $ {S}_k ^{\mathrm{T}}(t= 0) = 0$ corresponding to a
unexcited system with a fully filled valance band. The above equations
are equivalent to the semiconductor Bloch equations \cite{NLO} in the 
conduction-valence band representation with
Hamiltonian ${H}_{k,\mu\nu}$ and density
matrix ${\rho}_{k,\mu\nu}$ ($\mu,\nu = \pm$ denoting the conduction and
valence bands): 
the longitudinal and transverse
components of the spin density matrix are respectively related to the population
difference ${S}_k^{\mathrm{L}} ={\rho}_{k,++} - {\rho}_{k,--}$ and
interband coherence $ {S}_k^{\mathrm{T}} = -2i{\rho}_{k,+-}e^{i \omega
    t}$ between the two bands; whereas the longitudinal and transverse
components of the effective magnetic field are respectively related to the band
energy difference (including the intraband dipole matrix element) $\mathcal{B}_{k}^{\mathrm{L}} = 
{H}_{k,++}-{H}_{k,--}$ and interband dipole matrix element
$-2 i{H}_{k,+-}e^{i \omega
    t} = \mathcal{B}_{k}^{\mathrm{T}}$.  

\begin{figure}[!tb]
  \includegraphics[width=8.5cm,angle=0]{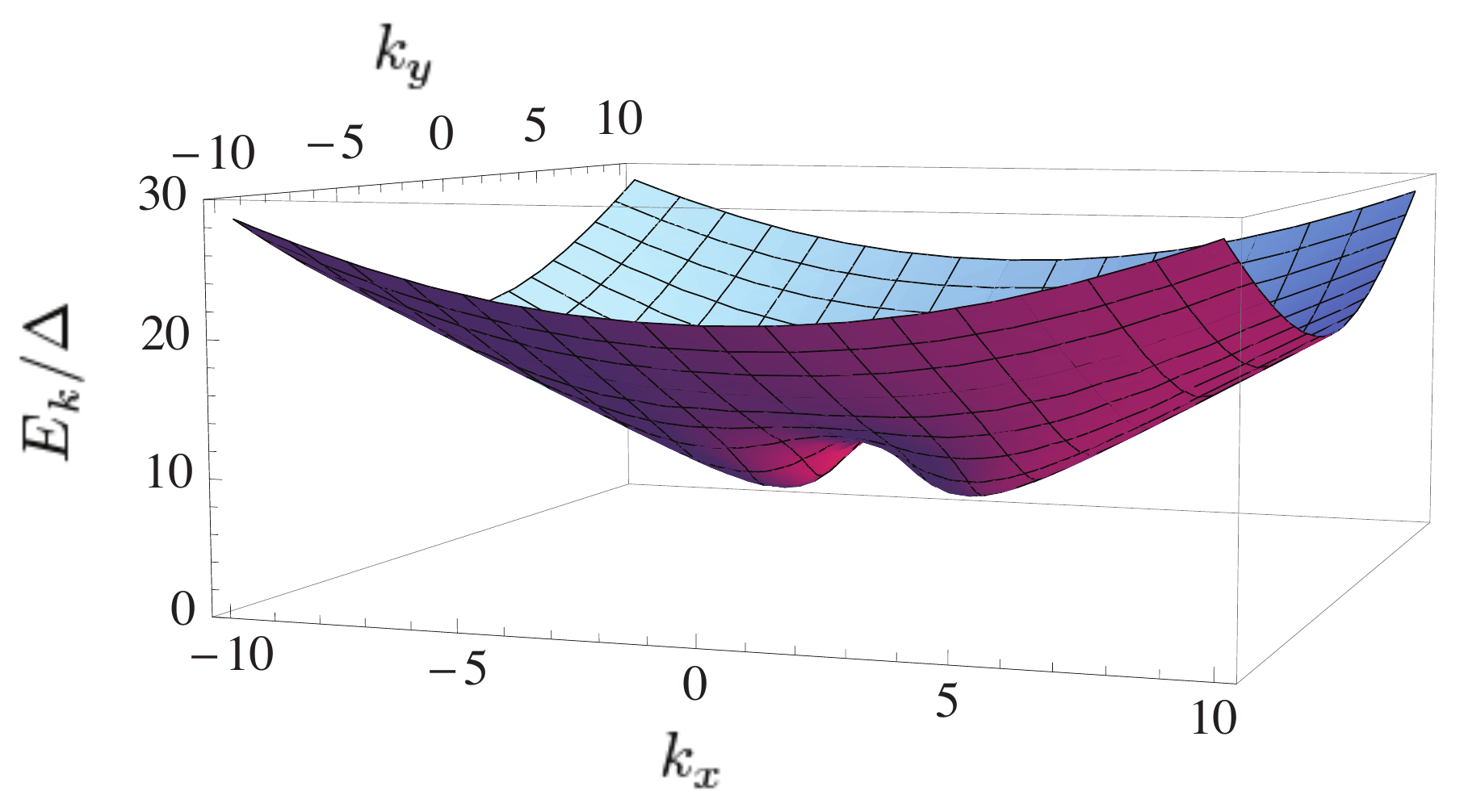}
\caption{(Color online). 3D plot of the renormalized conduction band
  energy $E_k$ (scaled by the gap $\Delta$) in the rotating frame. The
  surface Dirac gap $\Delta = 80\,\mathrm{meV}$, $\omega =
  60\,\mathrm{meV}$, $E = 150\,\mathrm{MVm}^{-1}$.} \label{RWABands}
\end{figure}

\begin{figure}[!tb]
  \includegraphics[width=8.5cm,angle=0]{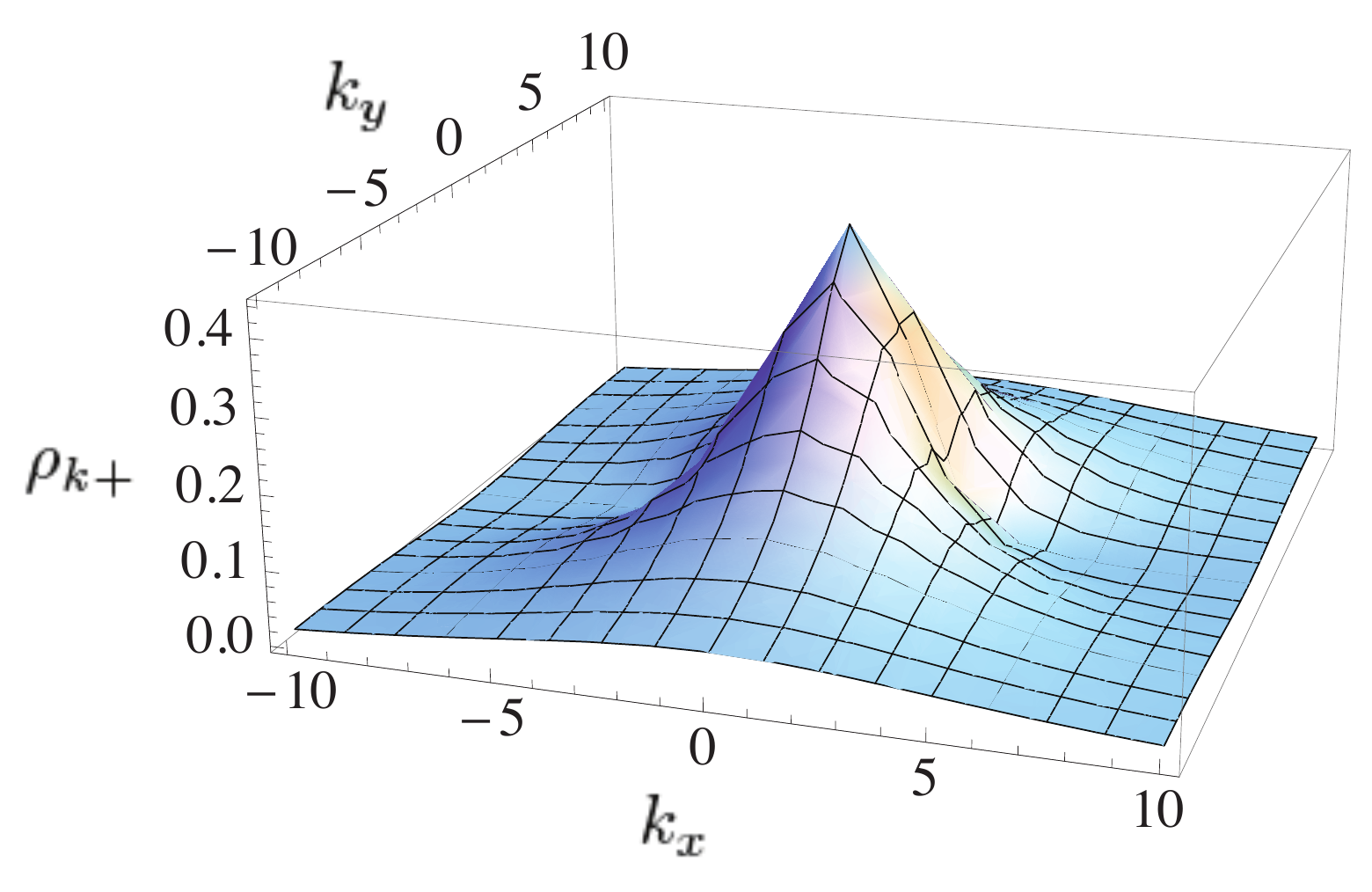}
\caption{(Color online). 3D plot of the conduction band distribution
  function $f_{k+}$ in the rotating frame. Values for $\Delta$ and $\omega$ are
  the same as in Fig.~\ref{RWABands}.} \label{RWADisf}
\end{figure}

We are interested in the optical response at the same frequency as
the incident field. Since the detuning $ \delta \ll 
  2\Delta$ is small, we work in the rotating wave approximation (RWA)
  keeping only the resonant terms. The
  longitudinal and transverse components of the effective magnetic
  field then becomes $\mathcal{B}_k^{\mathrm{L}} = 2\alpha_k-\omega$ and $\mathcal{B}_{k}^{\mathrm{T}} =
-\Lambda(\cos\theta_k\cos\phi_k+i\sin\phi_k)/2$. The quantity $\vert\mathcal{B}_{k}^{\mathrm{T}}\vert^2$,
corresponding to the squared amplitude of the interband transition
matrix element, appears frequently in our following 
discussions; for convenience, we shall denote the momentum-dependent factor
in $\vert\mathcal{B}_{k}^{\mathrm{T}}\vert^2$ as $X_{{k}}^2 =
\cos^2\theta_k\cos^2\phi_k+\sin^2\phi_k$. 
When the switch-on time of the laser pulse is much longer than 
$\Lambda^{-1}$, the switching process is adiabatic and 
the solution to Eqs. ~(\ref{SKE2})-(\ref{SKE3}) corresponds to the spin 
dynamics adiabatically following the effective magnetic field. 
Eqs.~(\ref{SKE2})-(\ref{SKE3}) then implies that 
$\left({S}_k^{\mathrm{T}}/\mathcal{B}_{k}^{\mathrm{T}}\right)^* =
\left({S}_k^{\mathrm{T}}/\mathcal{B}_{k}^{\mathrm{T}}\right) =
\left({S}_k^{\mathrm{L}}/\mathcal{B}_{k}^{\mathrm{L}}\right) \equiv
\mathcal{C}$, where $\mathcal{C}$ is a real constant. By virtue of the
normalization condition $\vert {S}_k^{\mathrm{L}}\vert^2+\vert
{S}_k^{\mathrm{T}}\vert^2 = 1$, the spin density matrix components are
found to be 
%
\begin{eqnarray}
{S}_k^{\mathrm{L}} &=& -\frac{\vert\mathcal{B}_{k}^{\mathrm{L}}\vert}{\sqrt{\vert \mathcal{B}_{k}^{\mathrm{L}}\vert^2 +
    \vert \mathcal{B}_{k}^{\mathrm{T}}\vert^2 }}, \label{SL} \\
{S}_k^{\mathrm{T}} &=& -\frac{\mathcal{B}_{k}^{\mathrm{T}}}{\sqrt{\vert\mathcal{B}_{k}^{\mathrm{L}}\vert^2+\vert\mathcal{B}_{k}^{\mathrm{T}}\vert^2}}. \label{ST}
\end{eqnarray}
The above distribution functions Eqs.~(\ref{SL})-(\ref{ST}) highlight the
semiconductor optical Stark effect \cite{Haug_OSE} under 
the adiabatic switch-on condition. 
Illuminated with a strong optical field, the
conduction and valence band states become mixed by the dipole matrix 
element with the system becoming a coherent ground state of photon dressed
electron-hole pairs. These dressed electron-hole pairs constitute the new
quasiparticles of the irradiated system with energy dispersions $\pm E_k =
\pm\sqrt{\vert\mathcal{B}_{k}^{\mathrm{L}}\vert^2+\vert\mathcal{B}_{k}^{\mathrm{T}}\vert^2}$
in the rotating frame. The leading-order energy shift is $\sim E_0^2$ 
characteristic of the optical Stark effect. 
From Eq.~(\ref{SL}), the conduction band distribution function is
found to be $\rho_{k+} = (1/2)(1+S_k^{\mathrm{L}})$. For the case of linearly
polarized light illumination on a 2D Dirac electron system, $\mathcal{B}_{k}^{\mathrm{T}}$ depends explicitly on
 $\phi_k$, and the renormalized band energies as well as 
the conduction and valance band distribution functions are anisotropic
in the momentum space as depicted in
Figs.~\ref{RWABands}-\ref{RWADisf}. Although light is off-resonance, in the
strong-field regime there is always a finite electron population in the conduction band due to nonlinear effects. 

\section{Dynamics of Current Response} \label{sec:current}

The current density in the system due to the \textit{A.C.} field is
given by $\bm{J}(t) = \sum_{k}\mathrm{tr}\{\bm{j}_k(t) f_k(t)\}$, where
$\bm{j}_k(t) = \partial H(t)/\partial \bm{k}$ is the single-particle current
operator. The current density can be written in terms of the effective magnetic field and the spin
density matrix as 
\begin{eqnarray}
\bm{J}(t) &=& \frac{1}{2}\sum_{k} \frac{\partial \bm{\mathcal{B}}_{k'}\cdot
  \bm{S}_k}{\partial \bm{k}'}\bigg\vert_{\bm{k}'=\bm{k}}. 
\label{J2}
\end{eqnarray}
Eqs.~(\ref{SL})-(\ref{ST}) are obtained in the rotating frame. 
In order to compute the current from Eq.~(\ref{J2}), we transform back into the
stationary frame. Then, the resonant contribution of the effective
magnetic field $\bm{\mathcal{R}}_k \equiv \bm{\mathcal{B}}_k^{{p}}\vert_{\mathrm{resonance}}$
due to
electron-photon coupling can be written as $\bm{\mathcal{R}}_k =
{\mathcal{R}}_k^{\beta}\hat{\boldsymbol{\beta}}_{k}+{\mathcal{R}}_k^{\gamma}\hat{\boldsymbol{\gamma}}_{k}$,
where ${\mathcal{R}}_k^{\beta} = \mathrm{Re}\left\{\mathcal{B}_k^{\mathrm{T}}e^{-i\omega
    t}\right\} = (\Lambda/2)(\cos\theta_k\cos\phi_k\cos\omega t
+\sin\phi_k\sin\omega t)$ and ${\mathcal{R}}_k^{\gamma} = -\mathrm{Im}\left\{\mathcal{B}_k^{\mathrm{T}}e^{-i\omega
    t}\right\} = (\Lambda/2)(\cos\theta_k\cos\phi_k\sin\omega t
  -\sin\phi_k\cos\omega t)$.  
%
%
The spin density matrix from Eqs.~(\ref{SL})-(\ref{ST}) is now expressed
in the stationary frame as 
%
\begin{eqnarray}
\bm{S}_k =
-f_k(\Lambda){\left[\left(2\alpha_k-\omega\right)
  \hat{\boldsymbol{\alpha}}_{k}+{\mathcal{R}}_k^{\beta}\hat{\boldsymbol{\beta}}_{k}+{\mathcal{R}}_k^{\gamma}\hat{\boldsymbol{\gamma}}_{k}\right]}, \label{SRWA}
\end{eqnarray}
where 
%
\begin{eqnarray}
f_k(\Lambda) = \frac{1}
{\sqrt{(2\alpha_k-\omega)^2+(\Lambda X_{{k}}/2)^2)}},
\label{f} 
\end{eqnarray}
where $X_{{k}}^2 = \cos^2\theta_k\cos^2\phi_k+\sin^2\phi_k$ as
defined in the paragraph before Eqs.~(\ref{SL})-(\ref{ST}). 
As usual, the sinuisoidal time dependence can be taken care of by defining
complex quantities associated with the exponential time factor
$e^{-i\omega t}$. We define the complex current density $\mathcal{J}$ through $\boldsymbol{J}(t) =
\mathrm{Re}\{\boldsymbol{\mathcal{J}}e^{-i\omega t}\}$. Using Eq.~(\ref{J2}), we find the longitudinal and Hall components
of the current density in response to the \textit{A.C.} field
\begin{eqnarray}
\mathcal{J}_x &=& -i\frac{e^2E_0v^2}{2}\sum_{{k}}\frac{f_{k}(\Lambda)}{\alpha_k}X_{{k}}^2,
\nonumber \\
\label{JxRWA} \\
\mathcal{J}_y &=& \frac{ e^2E_0v^2 \Delta }{2}\sum_{{k}} \frac{f_{k}(\Lambda)}{\alpha_k^2}. \label{JyRWA} 
\end{eqnarray}
%
%
%
%
%
%

\subsection{Weak Fields: Relation to Linear Response Regime} \label{sec:linear}
To see how the strong field regime is connected to the linear response
regime, it is instructive to expand the results
Eqs.~(\ref{JxRWA})-(\ref{JyRWA}) to leading order in powers of $E_0$. 
For 
weak fields such that $\Lambda
\ll  \delta$,  
we
expand Eqs.~(\ref{JxRWA})-(\ref{JyRWA}) in $\Lambda/\delta$ up to the leading
order. 
The integrals can be calculated analytically and yield 
\begin{eqnarray}
\mathcal{J}_x &=& -i\frac{e^2E_0}{8\pi}\left[\frac{\Delta^2}{\omega}\left(\frac{1}{\varepsilon_c}-\frac{1}{\Delta}\right)\right.
\label{JxRWA_E1}  \\
&&\left.-2\left(\frac{\Delta}{\omega}\right)^2\mathrm{ln}\bigg\vert\frac{\varepsilon_c}{\omega}\bigg\vert
  +
  \frac{1}{2}\left(1+4\frac{\Delta^2}{\omega^2}\right)\mathrm{ln}\bigg\vert
  \frac{\omega-2\varepsilon_c}{\omega-2\Delta}\bigg\vert\right], \nonumber \\
\mathcal{J}_y &=&  \frac{e^2E_0\Delta}{4\pi\omega}\left(\mathrm{ln}\bigg\vert\frac{\omega-2\varepsilon_c}{\varepsilon_c}\bigg\vert-\mathrm{ln}\bigg\vert\frac{\omega-2\Delta}{\Delta}\bigg\vert\right), 
\label{JyRWA_E1}  
\end{eqnarray}
%
%
%
where $\varepsilon_c$ is an ultraviolet energy cut-off that
corresponds the bandwidth of the surface Dirac bands, 
taken to be the bulk energy gap of the topological
insulator. We emphasize that the expressions for $\mathcal{J}_x$ and $\mathcal{J}_y$ above
are only valid for non-zero frequencies
$\omega \approx 2\Delta$. One cannot arrive at the
\textit{D.C.} limit by taking $\omega \to 0$ in the above expressions 
because the counter-rotating contributions, which are ignored in RWA, 
become comparable at $\omega \to 0$ to the rotating (\textit{i.e.},
resonant) contributions retained in the RWA. 
To recover the \textit{D.C.} limit, and indeed the full linear
response optical conductivity, one needs to add in the
counter-rotating contributons. 
To illustrate this point, we can explicitly take the $\omega \to 0$
limit and see what happens. First, $J_x(t)$ vanishes in this limit as
expected because $e^{-i\omega t} \to 1$ and Eq.~(\ref{JxRWA_E1}) is
purely imaginary. Then,  
expanding up to leading order in $\omega/\Delta$
in Eq.~(\ref{JyRWA_E1}), we find that $\mathcal{J}_y = e^2E_0/8\pi$
corresponding to a Hall conductivity of $\sigma_{xy} =
e^2/4h$. Interestingly, 
in the rotating-wave approximation where the counter-rotating 
contributions to the effective magnetic field are discarded, the
zero-frequency Hall response from Eq.~(\ref{JyRWA_E1}) amounts to
$1/2$ of the well-known quantized Hall
conductivity $\sigma_{xy} =
e^2/2h$ of the Dirac model in the linear response regime. 
Indeed, it can be easily checked that our weak-field results Eqs.~(\ref{JxRWA_E1})-(\ref{JyRWA_E1}) correspond exactly to
the resonant 
contribution of the established expressions of dynamical conductivities of the Dirac model
\cite{TI_MO1}. Adding in also the 
counter-rotating contribution, which is separately due to an effective magnetic
field $\bm{\mathcal{B}}_k^{p}\vert_{\mathrm{anti-resonance}} = \bm{\mathcal{R}}_k(-\omega)$
(noting that $\Lambda \to -\Lambda$ also under $\omega
\to -\omega$ in $\bm{\mathcal{R}}_k$ since $\Lambda$ is dependent on $\omega$),  
yields the full optical conductivities of
the Dirac model. 


\subsection{Strong Fields} \label{sec:strong}
For strong fields with $\Lambda \gg \delta$,  
we have 
\begin{equation}
f_k(\Lambda) \simeq \frac{2}{\Lambda}\frac{1}
{X_{{k}}}, 
\label{fapprox}
\end{equation}
Eqs.~(\ref{JxRWA})-(\ref{JyRWA}) can then be evaluated analytically
yielding 
\begin{eqnarray}
\mathcal{J}_x &=& -i\frac{e\omega}{2\pi^2 v
  \varepsilon_c} \label{JxRWA_str} \\
&&\times\left[\varepsilon_c^2
  \mathbb{E}\left(\sqrt{1-\frac{\Delta^2}{\varepsilon_c^2}}\right)-\Delta^2
  \mathbb{K}\left(\sqrt{1-\frac{\Delta^2}{\varepsilon_c^2}}\right)\right], \nonumber
\\
\mathcal{J}_y &=& \frac{e\omega\Delta}{2\pi^2 v }\left[\mathbb{K}\left(\sqrt{1-\frac{\Delta^2}{\varepsilon_c^2}}\right) 
-\mathbb{E}\left(\sqrt{1-\frac{\Delta^2}{\varepsilon_c^2}}\right)\right], 
\label{JyRWA_str}
\end{eqnarray}
where $\mathbb{K},\mathbb{E}$ are the complete elliptic integrals of
the first and second kind, respectively. At high fields, therefore, the longitudinal and Hall
currents both saturate to values independent of the incident field. 

\begin{figure}[!tb]
  \includegraphics[width=8.5cm,angle=0]{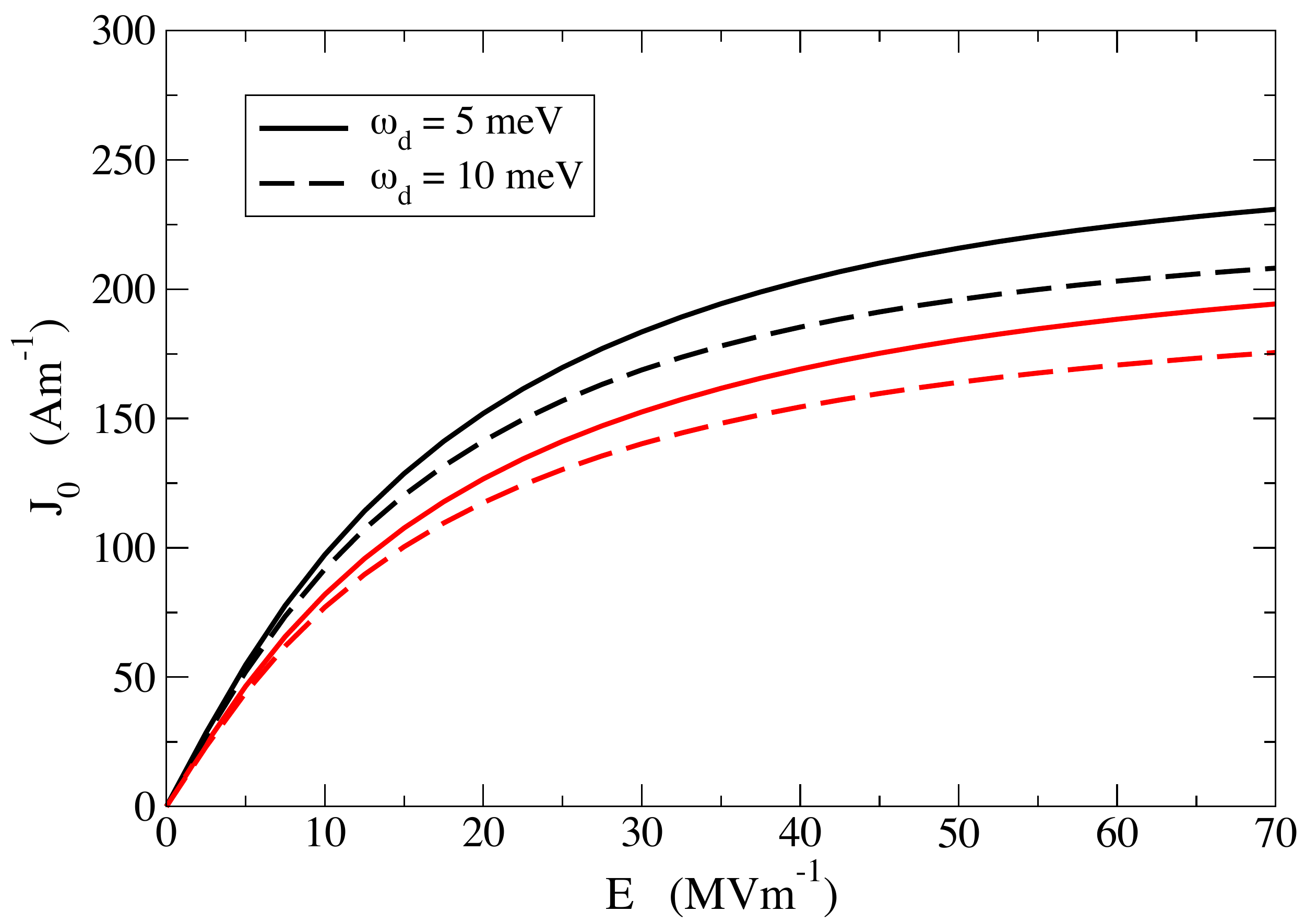}
\caption{(Color online) Amplitudes $ J_{x0}$ (dark/black) and $
  J_{y0}$ (grey/red) of the longitudinal and Hall current densities $J_{x0} =
  \vert \mathcal{J}_{x}\vert$ and $J_{y0} = \vert\mathcal{J}_{y}\vert$ as a function of the electric
  field $E_0$ [Eqs.~(\ref{JxRWA})-(\ref{JyRWA})] for different values of
  detuning $\omega_{\mathrm{d}} = 5\,\mathrm{meV}$ and
  $10\,\mathrm{meV}$. The surface Dirac gap $\Delta = 50\,\mathrm{meV}$ and
 Dirac band cutoff energy $\varepsilon_{\mathrm{c}} = 175\,\mathrm{meV}$.} \label{JRWA_detun}
\end{figure}

We evaluate Eqs.~(\ref{JxRWA})-(\ref{JyRWA}) numerically
and display the computed current densities in Fig.~\ref{JRWA_detun}. The
longitudinal and Hall currents increase linearly 
with the electric field and saturate at high $E_0$ values, as
predicted from the analytic results Eqs.~(\ref{JxRWA_E1})-(\ref{JyRWA_E1}) and Eqs.~(\ref{JxRWA_str})-(\ref{JyRWA_str}). 
We also see that increasing detuning has the effect of
decreasing the current amplitudes, consistent with the behavior that
the saturation currents are proportional to $\omega$ in
Eqs.~(\ref{JxRWA_str})-(\ref{JyRWA_str}).   

\section{Nonlinear Magneto-Optical Faraday and Kerr Effects} \label{sec:FaraKerr}



The usual Fresnel relations, which relate the transmitted and
reflected electric fields to the incident field, are derived assuming linear response to the incident light
field. Also, the scattering (or, equivalently, transfer) matrix formalism
assumes linearity between the scattered fields and the incident field. 
In order to take into account non-perturbative electric field effects
in the current response, these two standard 
approaches are not applicable and we need to formulate the problem of nonlinear
magneto-optical response differently as follows. 

Consider light illuminated along the $z$ direction
on the topological insulator surface that is located at $z = 0$. The
two regions $z < 0$ and $z > 0$ are labeled by $j = $I, II with dielectric constant $\epsilon_j$.
Expressing, as usual, the real electric and magnetic fields as complex
vectors $\boldsymbol{E} =
\mathrm{Re}\{\boldsymbol{\mathcal{E}}e^{-i\omega t}\}$ and $\boldsymbol{H} =
\mathrm{Re}\{\boldsymbol{\mathcal{H}}e^{-i\omega t}\}$ respectively, we write the electric 
field in the form 
\begin{eqnarray} 
\tilde{\mathcal{E}}^j = 
e^{ik_jz}\left[\begin{array}{c} \mathcal{E}_x^{tj} \\ \mathcal{E}_y^{tj}\end{array}\right]
+e^{-ik_jz}\left[\begin{array}{c} \mathcal{E}_x^{rj} \\ \mathcal{E}_y^{rj}\end{array}\right], \label{Ewave}
\end{eqnarray}
where the tilde accents denote column vectors $\tilde{\mathcal{E}} =
[\mathcal{E}_x\,\,\,\mathcal{E}_y]^{\mathrm{T}}$, the superscripts `$\mathrm{r}$' and `$\mathrm{t}$' on
$\mathcal{E}_{x,y}$ denote the reflected and transmitted field components, and $k_{j} = \sqrt{\epsilon_{j}}k_0$ is the
wavevector in region $j$ with dielectric constant $\epsilon_j$, $k_0 = \omega/c$ and $c$ is the speed of
light. The corresponding magnetic field is given by Faraday's law as 
\begin{eqnarray}
\tilde{\mathcal{H}}^j = \sqrt{{\epsilon_j}}\left\{
e^{ik_jz}\left[\begin{array}{c} -\mathcal{E}_y^{tj} \\
    \mathcal{E}_x^{tj} \end{array}\right] +e^{-ik_jz}\left[\begin{array}{c} \mathcal{E}_y^{rj} \\
-\mathcal{E}_x^{rj}\end{array}\right]\right\}, \label{Hwave}
\end{eqnarray}
The electric and magnetic fields at the interface $z = 0$ satisfy
the electrodynamic boundary conditions $\tilde{\mathcal{E}}^{\mathrm{I}} = \tilde{\mathcal{E}}^{\mathrm{II}}$
and $-i\tau_y(\tilde{\mathcal{H}}^{\mathrm{II}}-\tilde{\mathcal{H}}^{\mathrm{I}}) = (4\pi/c)\tilde{\mathcal{J}}$, 
where $(\tau_x,\tau_y,\tau_z)$ are Pauli matrices and $\tilde{\mathcal{J}} =
[\mathcal{J}_x\,\,\,\mathcal{J}_y]^{\mathrm{T}}$ is the current density of
the topological surface states at $z = 0$. 

We calculate the transmission and reflection coefficients due to 
incident light that is linearly polarized along $x$, $\tilde{\mathcal{E}}^{\mathrm{I}} = 
e^{ik_0z}[E_0\,\,\,0]^{\mathrm{T}}$. For a single interface,  the
scattered field components in region I and region II correspond to the
reflected and transmitted fields respectively, and  to simplify the
notation we shall drop the
superscripts I and II with no danger of confusion.
The transmission and reflection coefficents along the directions $\alpha = x, y$ are defined as $T_{\alpha} =
\mathcal{E}_{\alpha}^t/E_0$ and $R_{\alpha} = \mathcal{E}_{\alpha}^r/E_0$, which are functions of $E_0$ in the nonlinear regime. From the
electromagnetic boundary conditions we then obtain the following set
of coupled nonlinear equations for $T_{x}$ and $T_y$:
\begin{eqnarray}
T_x &=&
1+\pi\alpha v^2\sum_{{k}}\frac{1}{\alpha_k} \label{Tx} \\
&&\times\left\{f_k(T_y^2\Lambda^2)\frac{\Delta}{\alpha_k}T_y +i
  f_k(T_x^2\Lambda^2)X_{{k}}^2T_x\right\},  \nonumber \\
T_y &=& -\pi\alpha v^2\sum_{{k}}\frac{1}{\alpha_k} \label{Ty} \\
&&\times\left\{f_k(T_x^2\Lambda^2)\frac{\Delta}{\alpha_k}T_x-i
  f_k(T_y^2\Lambda^2)X_{{k}}^2T_y\right\}, \nonumber 
\end{eqnarray}
where 
$\alpha = e^2/\hbar c$ is the
fine structure constant. The reflection coefficients are 
related to the transmission coefficients as $R_x = T_x-1$ and $R_y =
T_y$. In the linear regime, these equations can be decoupled easily
and reduce to the 
familar
relations \cite{TI_MO3} for the transmission coefficients
%
\begin{eqnarray}
T_x &=&
\frac{1+2\pi\sigma_{xx}/c}{\left(1+2\pi\sigma_{xx}/c\right)^2+\left(2\pi\sigma_{yx}/c\right)^2}, \label{TxLin} \\
T_y &=&
-\frac{2\pi\sigma_{yx}/c}{\left(1+2\pi\sigma_{xx}/c\right)^2+\left(2\pi\sigma_{yx}/c\right)^2}, \label{TyLin}
\end{eqnarray}
where $\sigma_{xx}$ and $\sigma_{yx}$ are respectively the longitudinal and Hall
conductivities under RWA, given by
Eqs.~(\ref{JxRWA_E1})-(\ref{JyRWA_E1}) through $\sigma_{xx} =
\mathcal{J}_x/E_0$ and $\sigma_{yx} = \mathcal{J}_y/E_0$. 
In the strong field regime, Eqs.~(\ref{Tx})-(\ref{Ty}) 
must be solved simultaneously in order to obtain the
transmission coefficients $T_{x,y}$.  
%
%
The Faraday $\theta_F$ and Kerr $\theta_K$ angles are connected in 
the usual way to the transmission and reflection coefficients as $\theta_F = \left[\mathrm{arg}(T_{-})-\mathrm{arg}(T_{+})\right]/2$ and
$\theta_K = \left[\mathrm{arg}(R_{-})-\mathrm{arg}(R_{+})\right]/2$, where
`$\mathrm{arg}$' denotes taking the complex argument, $T_{\pm} =
T_x\pm iT_y$ and $R_{\pm} =
R_x\pm iR_y$ stand for the transmission and reflection coefficients for the
$\pm$ circularly polarized components of the transmitted and reflected
light, respectively. 

\begin{figure}[!tb]
  \includegraphics[width=8.5cm,angle=0]{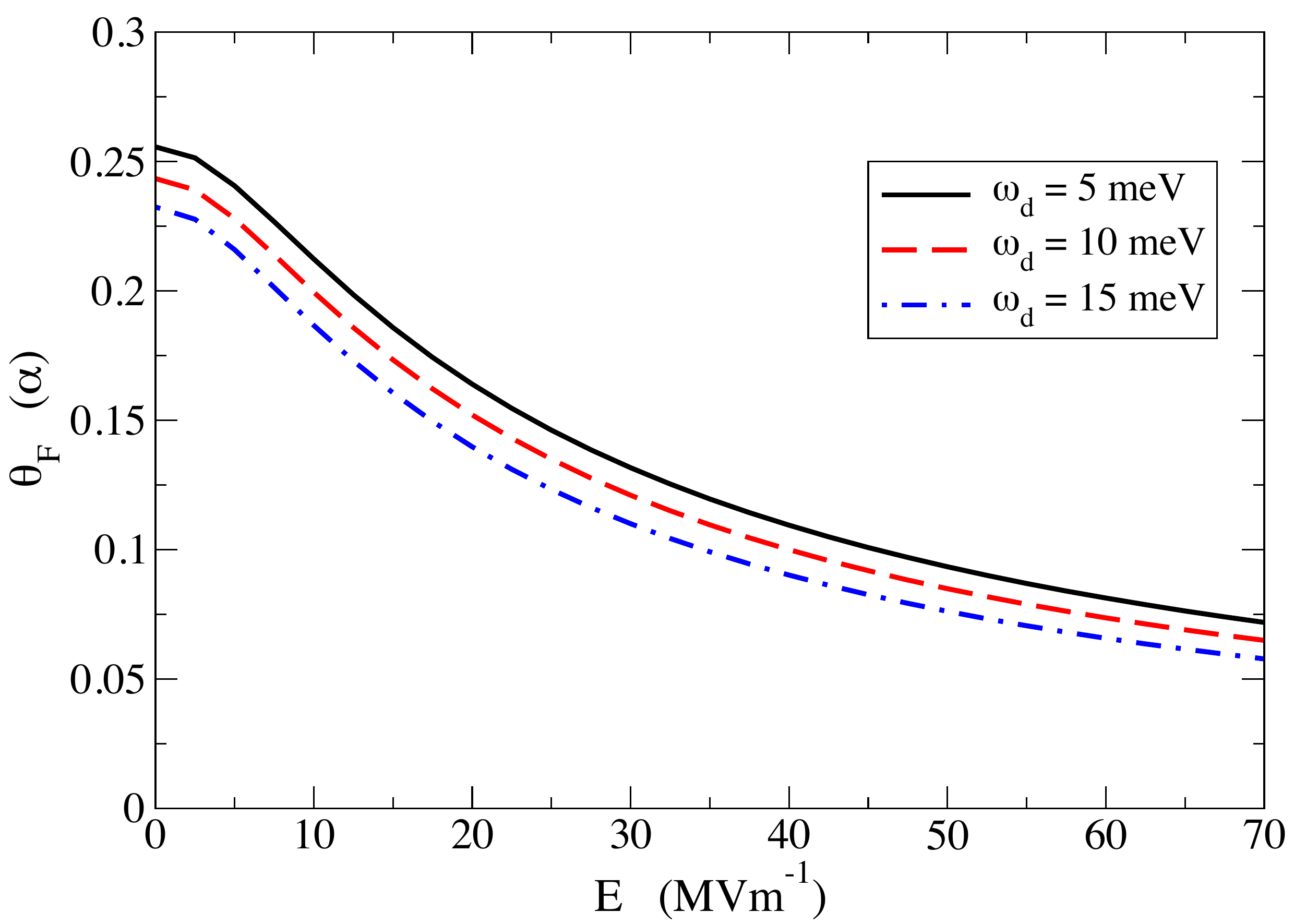}
\caption{(Color online) Faraday angle $\theta_F$ (in units of $\alpha$
  radians, where $\alpha = 1/137$ is the fine
  structure constant) versus electric 
  field $E_0$ for different values of
  detuning $\omega_{\mathrm{d}} = 5\,\mathrm{meV}$, 
  $10\,\mathrm{meV}$,  $15\,\mathrm{meV}$. The parameters $\Delta$ and
$\varepsilon_{\mathrm{c}}$ are the same as that in
Fig.~\ref{JRWA_detun}.} \label{FaraRWA_detun}
\end{figure}

\begin{figure}[!tb]
  \includegraphics[width=8.5cm,angle=0]{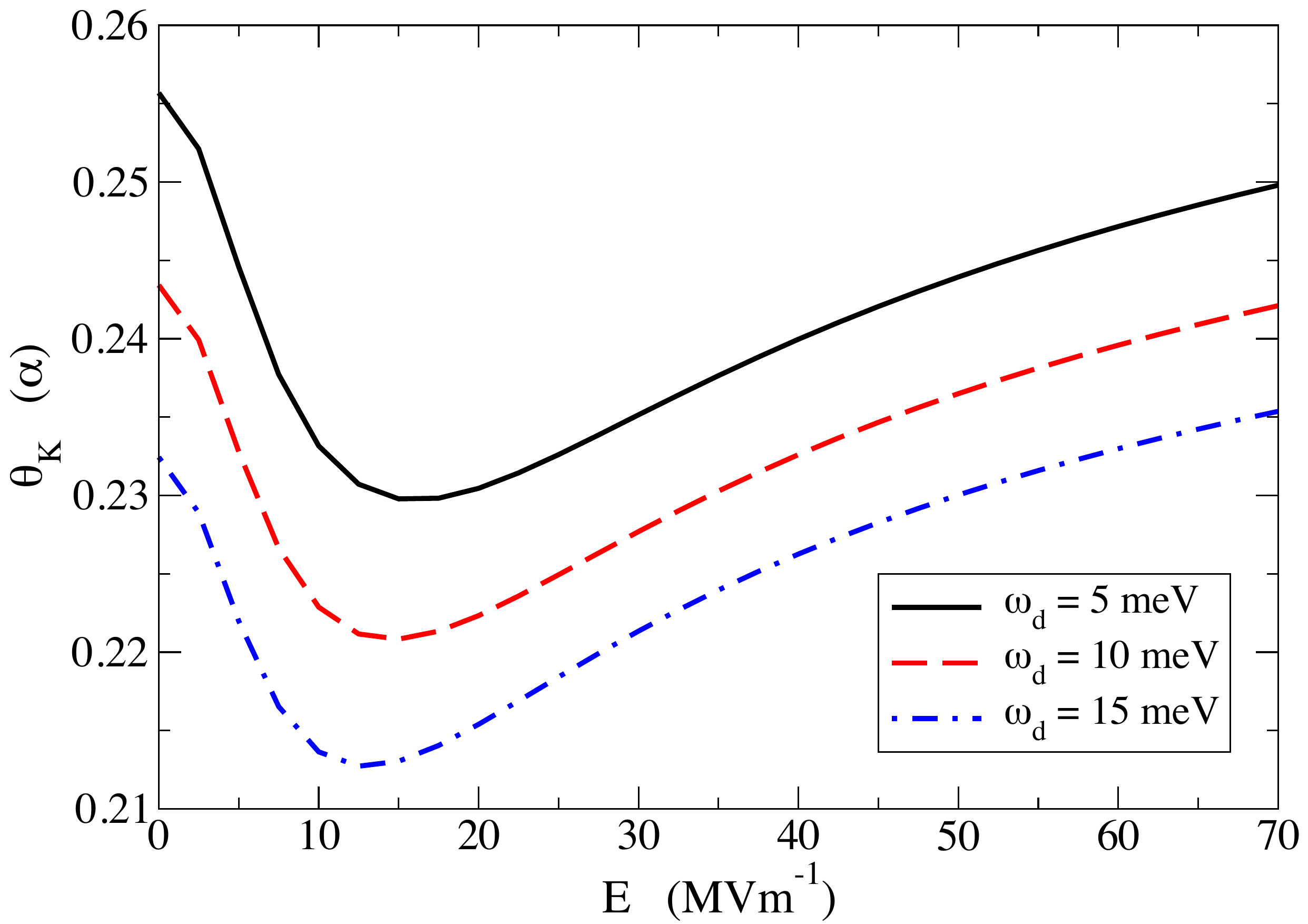}
\caption{(Color online) Kerr angle $\theta_K$ (in units of
  $\alpha$ radians) versus electric
  field $E_0$ for different values of
  detuning $\omega_{\mathrm{d}} = 5\,\mathrm{meV}$, 
  $10\,\mathrm{meV}$,  $15\,\mathrm{meV}$. 
} \label{KerrRWA_detun}
\end{figure}

Figs.~\ref{FaraRWA_detun}-\ref{KerrRWA_detun} show the Faraday and
Kerr angles calculated from numerically solving
Eqs.~(\ref{Tx})-(\ref{Ty}). The values approaching $E_0 = 0$
correspond to the results from weak-field regime where $T_{x,y}$ are
independent of the incident field. One might naively expect that both angles
increase with the electric field. On the contrary, we find that $\theta_{F}$ decreases with the incident electric
field whereas $\theta_K$ exhibits an interesting non-monotonic trend,
decreasing first before going through an upturn. 
The leading-order nonlinear corrections for both effects are second
order, a signature of optical Stark effect, going as
$\theta_{F,K} \sim  A-B E_0^2$ where $A,B$ are positive
constants.  
We also note that both  $\theta_{F}$ and $\theta_{K}$ increases
with decreasing detuning. 


\section{Effects of Relaxation} \label{sec:relax}

We now consider the regime when the frequency is close to the
surface band gap with the detuning smaller than or on the same order of
magnitude as the inverse lifetime $1/\tau$ of the conduction
electrons. To more fully characterize lifetime effects, 
we introduce the phenomenological longitudinal $\Gamma$ and transverse $\Gamma_{\perp}$ relaxation rates
into the Bloch equation employing a relaxation time approximation. $\Gamma$ relaxes the conduction band 
electrons to the valence band due to momentum-independent processes  
such as radiative recombination and optical phonon scattering, whereas 
$\Gamma_{\perp}$ accounts for the decoherence effects on the 
polarization. Eq.~(\ref{SKE2}) then becomes 
\begin{eqnarray}
i\frac{\partial  {S}_k ^{\mathrm{L}}}{\partial t} &=& 
\frac{1}{2}\left[\left({S}_k
    ^{\mathrm{T}}\right)^*\mathcal{B}_{k}^{\mathrm{T}}-{S}_k
    ^{\mathrm{T}}\left(\mathcal{B}_{k}^{\mathrm{T}}\right)^*\right]-i\Gamma\left({S}_k
  ^{\mathrm{L}}+1\right), \nonumber \\ \label{SKEG1}
\\
i\frac{\partial  {S}_k ^{\mathrm{T}}}{\partial t} &=&
\frac{1}{2}\left[{S}_k^{\mathrm{L}}\left(\mathcal{B}_{k}^{\mathrm{T}}\right)^*-\left({S}_k^{\mathrm{T}}\right)^*\mathcal{B}_{k}^{\mathrm{L}}\right]-i\Gamma_{\perp}{S}_k ^{\mathrm{T}}. \label{SKEG2}
\end{eqnarray}

With damping, the spin density matrix no
longer satisfy the unitarity condition $\vert \bm{S}_k \vert = 1$
since the number of electrons in the system is not conserved. 
After the light field has been turned on and the transients
have subsided, $\bm{S}_k$
approaches a steady-state value dependent on its initial state 
and the damping parameters $\Gamma,\Gamma_{\perp}$. 
The steady-state solution is obtained in the rotating frame by
requiring, within RWA, that ${\partial  {S}_k
  ^{\mathrm{L,T}}}/{\partial t} = 0$.  
We then tranform the resulting
expressions back into the stationary frame 
in order to obtain the currents from Eq.~(\ref{J2}), yielding  
%
\begin{eqnarray}
\bm{S}_k &=& -g_k(\Lambda)\left\{\left[\left(2\alpha_k-\omega\right)^2+\Gamma_{\perp}^2\right]\hat{\boldsymbol{\alpha}}_{k}\right.
\label{SRWAG} \\
&&+\left[\left(2\alpha_k-\omega\right)\mathcal{R}_k^{\beta}+\Gamma_{\perp}\mathcal{R}_k^{\gamma}\right]\hat{\boldsymbol{\beta}}_{k}
\nonumber \\
&&\left.+\left[\left(2\alpha_k-\omega\right)\mathcal{R}_k^{\gamma}-\Gamma_{\perp}\mathcal{R}_k^{\beta}\right]\hat{\boldsymbol{\gamma}}_{k}\right\}, \nonumber
\end{eqnarray}
where 
\begin{eqnarray}
g_{k}(\Lambda) =  
\frac{1}{(2\alpha_k-\omega)^2+\Gamma_{\perp}^2+\left({\Gamma_{\perp}}/{\Gamma}\right)\left({\Lambda
      X_{{k}}}/{2}\right)^2}. \label{g} 
\end{eqnarray}
%
Eqs.~(\ref{SRWAG})-(\ref{g}) are valid in the steady-state regime for nonzero
$\Gamma,\Gamma_{\perp}$. 
Using Eq.~(\ref{J2}), we then find the longitudinal and Hall current densities 
\begin{eqnarray}
\mathcal{J}_x &=& i\frac{e^2E_0v^2}{2\omega}\sum_k
\frac{g_{k}(\Lambda)}{\alpha_k}X_{{k}}^2\left(\omega^2+\Gamma_{\perp}^2-2\alpha_k\omega-i2\Gamma_{\perp}\alpha_k\right), \nonumber \\
\label{JxRWAG} \\
\mathcal{J}_y &=& -\frac{e^2E_0v^2\Delta}{2\omega}\sum_k
\frac{g_{k}(\Lambda)}{\alpha_k^2}
\left(\omega^2+\Gamma_{\perp}^2-2\alpha_k\omega-i2\Gamma_{\perp}\alpha_k\right). \nonumber
\\ 
\label{JyRWAG} 
\end{eqnarray}
%
%
%
The presence of relaxation and dephasing introduces dissipation in the 
current response. In Eqs.~(\ref{JxRWAG})-(\ref{JyRWAG}), 
$\mathrm{Im}\left\{\mathcal{J}_x\right\}$ and
$\mathrm{Re}\left\{\mathcal{J}_y\right\}$ correspond to reactive, 
non-dissipative current components whereas $\mathrm{Re}\left\{\mathcal{J}_x\right\}$ and
$\mathrm{Im}\left\{\mathcal{J}_y\right\}$ correspond to dissipative
components. 

\subsection{Weak Fields} \label{sec:linearG}
For $\Lambda \ll \delta \lesssim  \Gamma,\Gamma_{\perp}$, we expand $\mathcal{J}_{x,y}$ in 
Eqs.~(\ref{JxRWAG})-(\ref{JyRWAG}) to leading order in
$\Lambda/\delta$ and obtain
%
\begin{eqnarray}
\mathcal{J}_x &=& -\frac{e^2E_0}{16 \pi\omega}
\left\{\left(\omega+i\Gamma_{\perp}+\frac{4\Delta^2}{\omega+i\Gamma_{\perp}}\right)\right. \nonumber\\
&&\times\left[\tan^{-1}\left(\frac{\omega-2\varepsilon_c}{\Gamma_{\perp}}\right)-\tan^{-1}\left(\frac{\omega-2\Delta}{\Gamma_{\perp}}\right)\right.
  \nonumber \\
&&\left.+\frac{i}{2}\mathrm{ln}\left[\frac{\left(\omega-2\varepsilon_c\right)^2+\Gamma_{\perp}^2}{\left(\omega-2\Delta\right)^2+\Gamma_{\perp}^2}\right]\right]
  \nonumber \\
&&\left.+i2\Delta^2\left[\frac{1}{\varepsilon_c}-\frac{1}{\Delta}-\frac{2}{\omega+i\Gamma_{\perp}}\mathrm{ln}\left(\frac{\varepsilon_c}{\Delta}\right)\right]\right\},
\label{JxG} 
\end{eqnarray}
\begin{eqnarray}
\mathcal{J}_y &=& 
-\frac{e^2E_0\Delta}{4\pi\omega}\left\{\mathrm{ln}\left(\frac{\varepsilon_c}{\Delta}\right)-\frac{1}{2}\mathrm{ln}\left[\frac{\left(\omega-2\varepsilon_c\right)^2+\Gamma_{\perp}^2}{\left(\omega-2\Delta\right)^2+\Gamma_{\perp}^2}\right] \right.
\nonumber \\
&&\left.+i\left[\tan^{-1}\left(\frac{\omega-2\varepsilon_c}{\Gamma_{\perp}}\right)-\tan^{-1}\left(\frac{\omega-2\Delta}{\Gamma_{\perp}}\right)\right]\right\}. \nonumber
\\ 
\label{JyG}
\end{eqnarray}

We note that the longitudinal relaxation rate 
$\Gamma$ does not come into the expressions Eqs.~(\ref{JxG})-(\ref{JyG}) for
$\mathcal{J}_{x,y}$. In the weak-field regime, therefore, relaxation
processes such as radiative recombination and electron-optical phonon scattering that
relax conduction band electrons back into the valence band do not
contribute. This is because these processes only happen when there are
real transitions resulting in a finite electron population in the
conduction band. For subgap frequencies, a finite electron population
in the conduction band only occurs through nonlinear correction $\sim
E_0^2$, and are absent in the linear $\sim E_0$ regime. 

Despite $\Gamma,\Gamma_{\perp} > 0$ in the regime considered in this
section, we note that 
the $\Gamma_{\perp} \to 0$ limit of Eqs.~(\ref{JxG})-(\ref{JyG})
recovers Eqs.~(\ref{JxRWA_E1})-(\ref{JyRWA_E1}) obtained in the
coherent regime where the effects of damping are ignored.
This correspondence however is restricted only to the linear regime,
and does not hold when nonlinear effects in the electric field come in. 
 

\subsection{Strong Fields} \label{sec:strongG}

For strong fields $\Lambda \gg \delta, \Gamma, \Gamma_{\perp}$, we
have 
\begin{equation}
g_k(\Lambda) \simeq  \frac{4 \Gamma}{\Gamma_{\perp}\Lambda^2}\frac{1}{X_{{k}}^2}.
\label{gapprox} 
\end{equation}
Eqs.~(\ref{JxRWAG})-(\ref{JyRWAG}) can then be evaluated analytically
yielding 
\begin{eqnarray}
\mathcal{J}_x &=& \frac{i\omega}{ 4\pi E_0
  v^2}\frac{\Gamma}{\Gamma_{\perp}}\left(\varepsilon_c-\Delta\right)\left(\omega+i\Gamma_{\perp}\right)
\nonumber \\
&&\times\left(\omega-i\Gamma_{\perp}-\varepsilon_c-\Delta\right) \label{JxRWAG_str} \\
\mathcal{J}_y &=& -\frac{\omega}{4\pi E_0
  v^2}\frac{\Gamma}{\Gamma_{\perp}}\left(\varepsilon_c-\Delta\right)\left(\omega+i\Gamma_{\perp}\right) \nonumber \\
&&\times\left(\omega-i\Gamma_{\perp}-\varepsilon_c-\Delta\right) \label{JyRWAG_str} 
\end{eqnarray}
%

\begin{figure}[!tb]
  \includegraphics[width=8.5cm,angle=0]{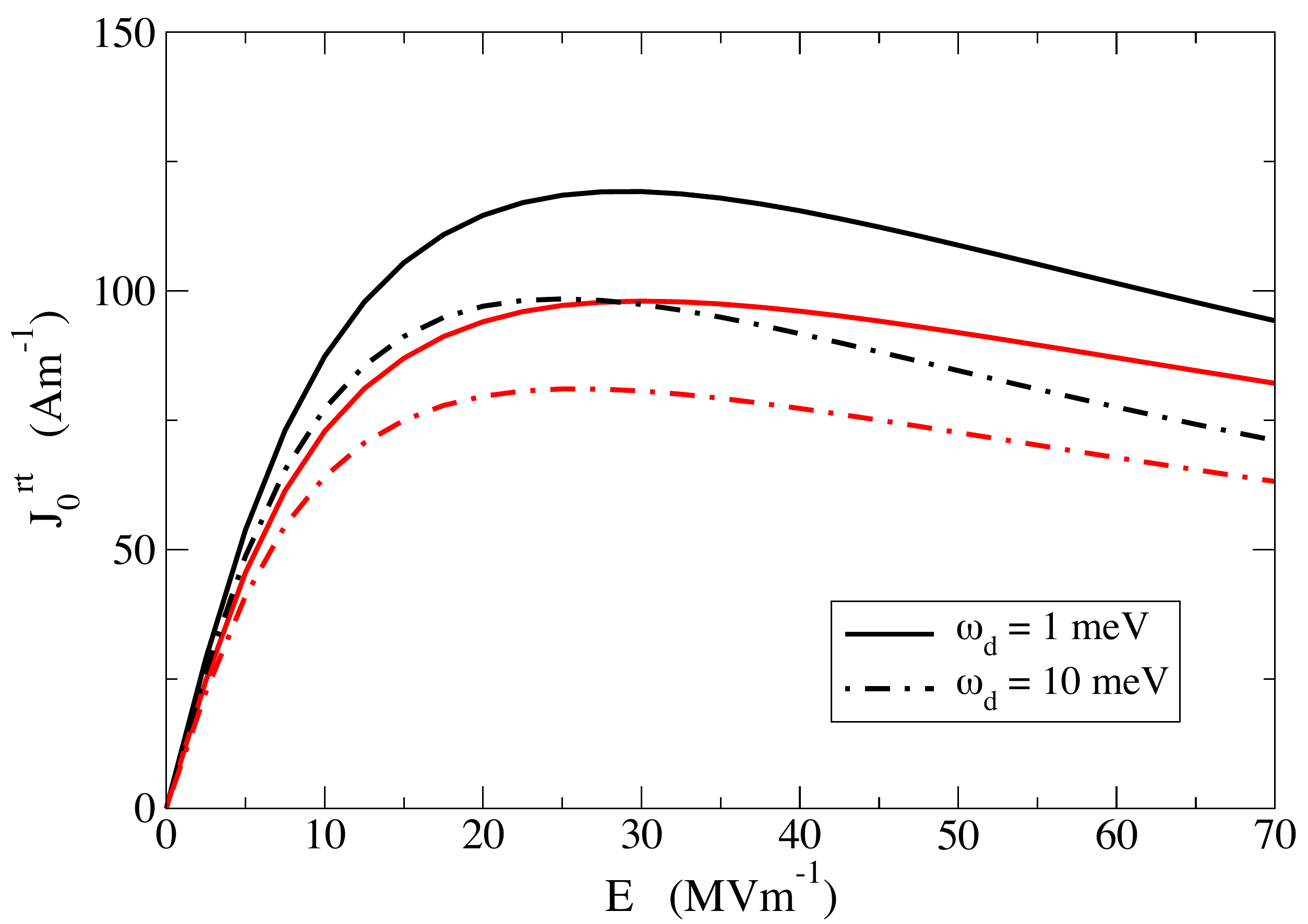}
\caption{(Color online) Amplitudes of the reactive components   (denoted by the superscript 'rt')
  $J_{x0}^{\mathrm{rt}} = \mathrm{Im}\left\{\mathcal{J}_x\right\}$ (dark/black) and $
  J_{y0}^{\mathrm{rt}} = \mathrm{Re}\left\{\mathcal{J}_y\right\}$ (grey/red) of the longitudinal and Hall
  current densities, as a function of the electric
  field $E_0$ for $\Gamma = \Gamma_{\perp} = 5\,\mathrm{meV}$ and different values of
  detuning $\omega_{\mathrm{d}} = 1\,\mathrm{meV}$ and
  $10\,\mathrm{meV}$. The values of the surface Dirac gap and
 Dirac band cutoff energy are the same as in Fig.~\ref{JRWA_detun}.} \label{JrtRWA_G_detun}
\end{figure}

\begin{figure}[!tb]
  \includegraphics[width=8.5cm,angle=0]{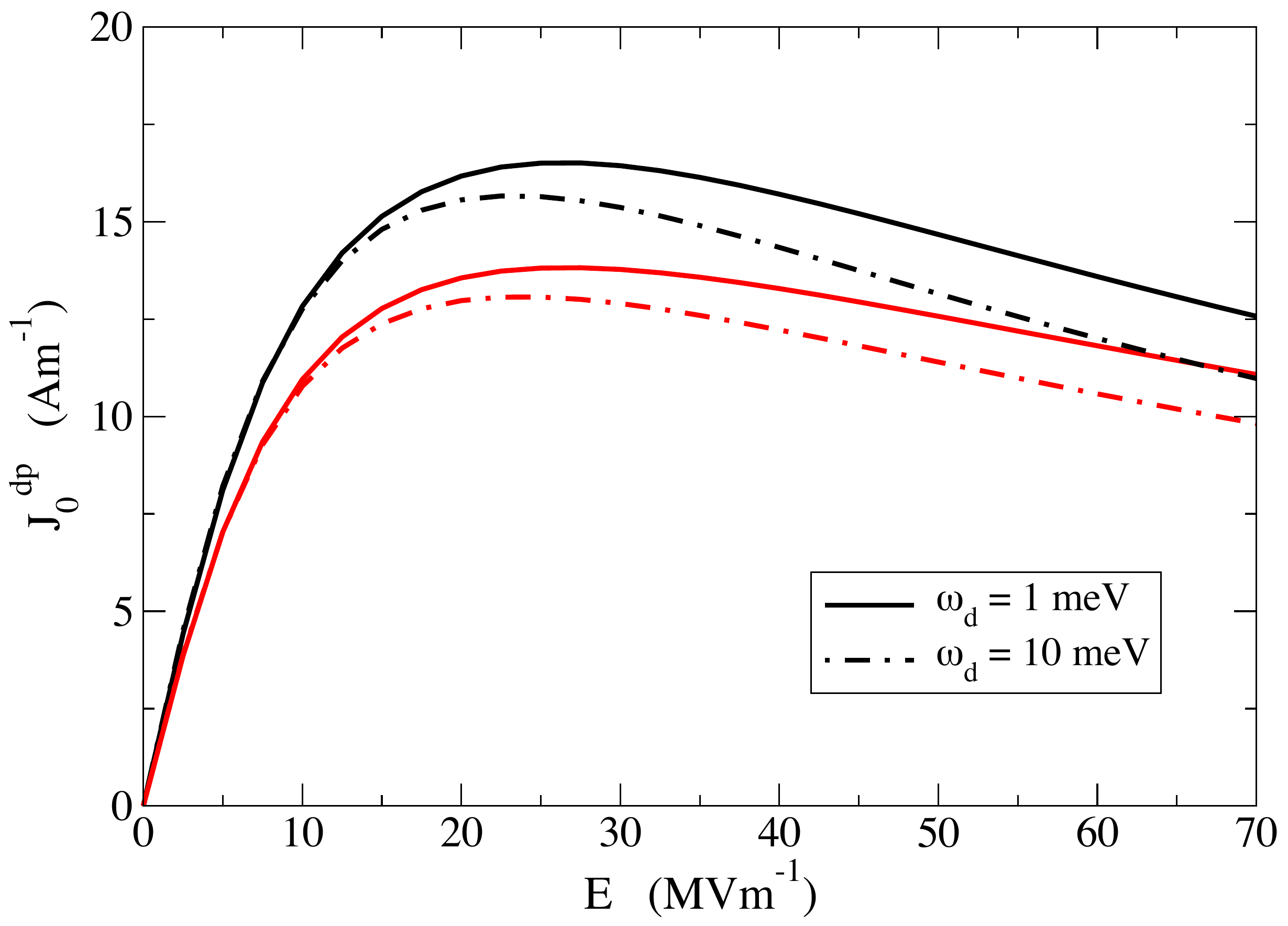}
\caption{(Color online) Amplitudes of the dissipative components
  (denoted by the superscript 'dp')
  $J_{x0}^{\mathrm{dp}} = \mathrm{Re}\left\{\mathcal{J}_{x}^{\mathrm{dp}}\right\}$ (dark/black) and $J_{y0}^{\mathrm{dp}} =
 \mathrm{Im}\left\{\mathcal{J}_{y}^{\mathrm{dp}}\right\}$ (grey/red) of the longitudinal and Hall
  current densities, as a function of the electric
  field $E_0$ for $\Gamma = \Gamma_{\perp} = 5\,\mathrm{meV}$ and different values of
  detuning $\omega_{\mathrm{d}} = 1\,\mathrm{meV}$ and
  $10\,\mathrm{meV}$. } \label{JdpRWA_G_detun}
\end{figure}

In contrast to the results we find in the coherent regime where both the 
longitudinal and Hall currents saturate at high $E_0$ values, in our 
present case we find that $\mathcal{J}_{x,y}$ do not saturate but instead
decrease at high electric fields as $1/E_0$. 
Fig.~\ref{JrtRWA_G_detun} shows the amplitudes of the reactive components of the
longitudinal and Hall current densities calculated from 
  Eqs.~(\ref{JxRWAG})-(\ref{JyRWAG}). Comparing the results for
  $\omega_{\mathrm{d}} = 10\,\mathrm{meV}$ in the absence and presence
  of $\Gamma,\Gamma_{\perp}$ (Fig.~\ref{JRWA_detun} and
  Fig.~\ref{JrtRWA_G_detun}, respectively), we see that
  relaxation and dephasing decrease the reactive components of
  the longitudinal and Hall currents. In the regime considered where $\delta \lesssim \Gamma,
  \Gamma_{\perp}$, the dissipative 
  current components are finite but still small, about an order of magnitude less than that of the reactive
  components (Fig.~\ref{JdpRWA_G_detun}). 

%

\subsection{Nonlinear Faraday and Kerr Effects} \label{sec:FaraKerrG}

Including relaxation and dephasing rates, we find that the coupled
equations for $T_x$ and $T_y$ are modified as follows 
\begin{eqnarray}
T_x &=&
1-\frac{\pi\alpha
  v^2}{\omega}\sum_{\bm{k}}\frac{\omega^2+\Gamma_{\perp}^2-2\alpha_k\omega-i2\Gamma_{\perp}\alpha_k}{\alpha_k}
\label{TxG} \\
&&\times\left\{g_k(T_y^2\Lambda^2)\frac{\Delta}{\alpha_k}T_y
+i g_k(T_x^2\Lambda^2)X_{{k}}^2T_x\right\}, \nonumber \\
T_y &=& \frac{\pi\alpha
  v^2}{\omega}\sum_{\bm{k}}\frac{\omega^2+\Gamma_{\perp}^2-2\alpha_k\omega-i2\Gamma_{\perp}\alpha_k}{\alpha_k}
\label{TyG} \\
&&\times\left\{g_k(T_x^2\Lambda^2)\frac{\Delta}{\alpha_k}T_x -i
  g_k(T_y^2\Lambda^2)X_{{k}}^2T_y\right\}. \nonumber
\end{eqnarray}
\begin{figure}[!tb]
  \includegraphics[width=8.5cm,angle=0]{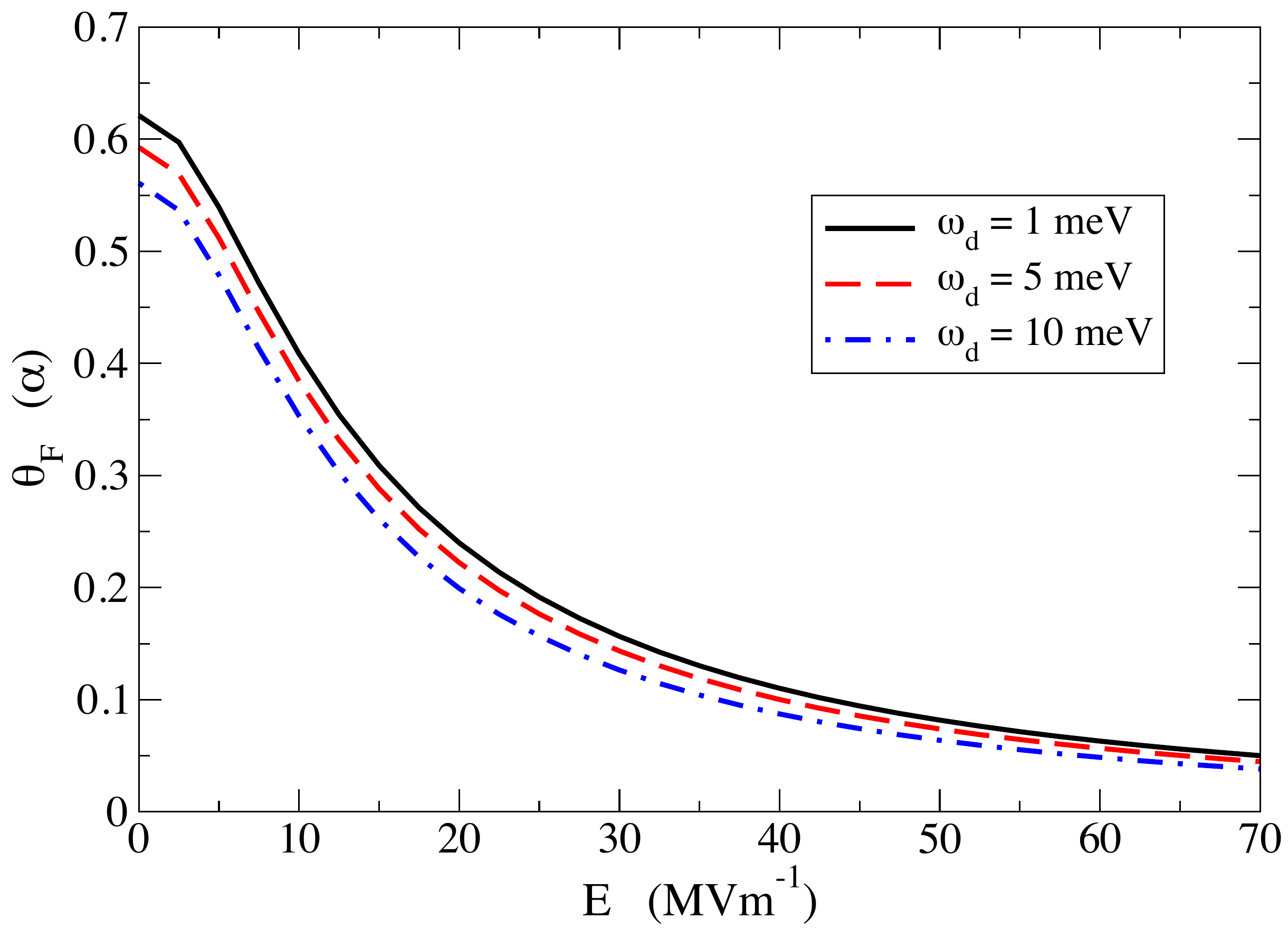}
\caption{(Color online) Faraday angle $\theta_F$ (in units of
  $\alpha$ radians) versus electric
  field $E_0$ for $\Gamma = \Gamma_{\perp} = 5\,\mathrm{meV}$ and different values of
  detuning $\omega_{\mathrm{d}} = 1\,\mathrm{meV}$, 
  $5\,\mathrm{meV}$,  $10\,\mathrm{meV}$. The parameters $\Delta$ and
$\varepsilon_{\mathrm{c}}$ are the same as that in Fig.~\ref{JRWA_detun}.} \label{FaraRWA_Gamma}
\end{figure}
\begin{figure}[!tb]
  \includegraphics[width=8.5cm,angle=0]{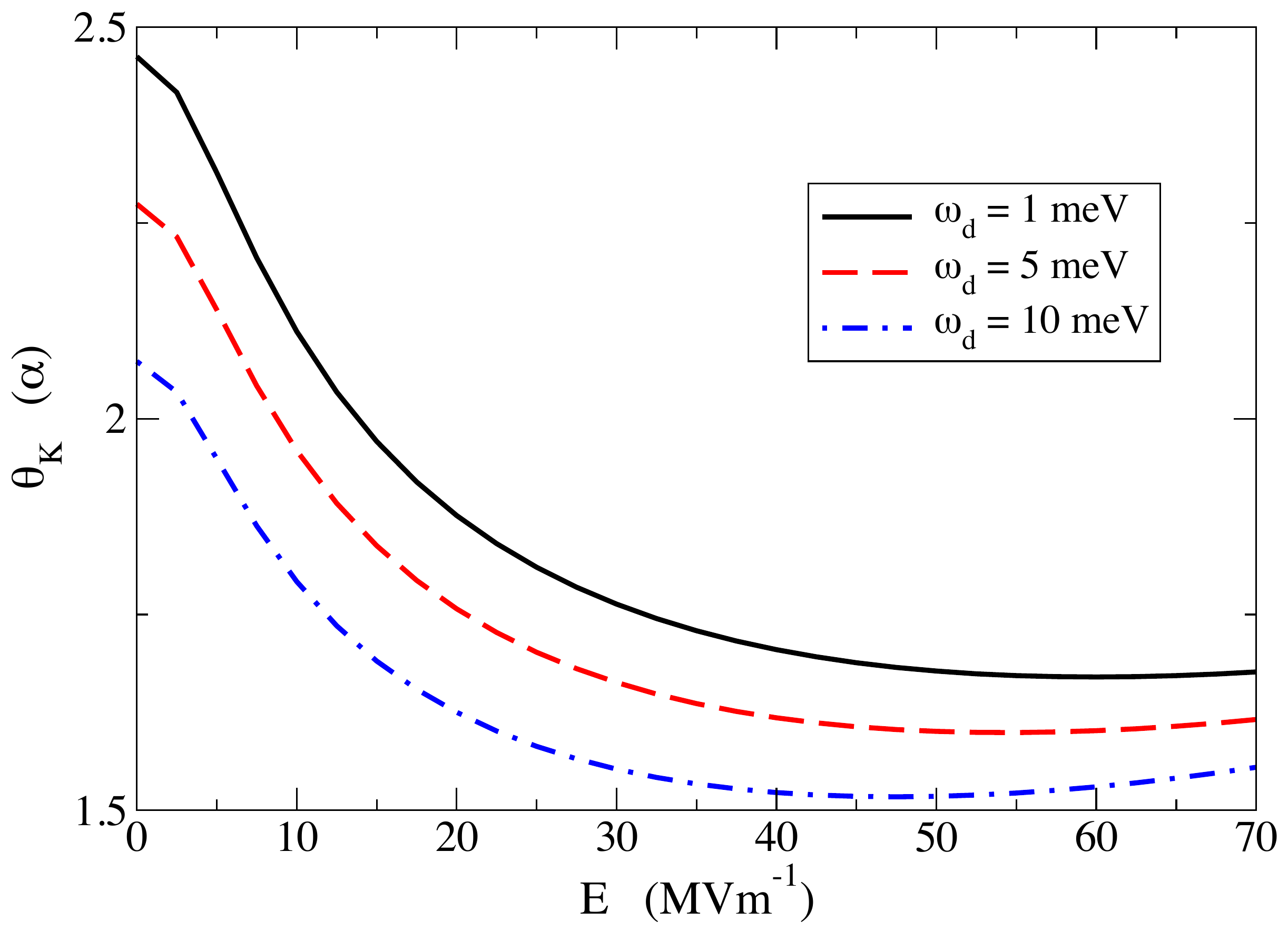}
\caption{(Color online) Kerr angle $\theta_K$ (in units of
  $\alpha$ radians) versus electric
  field $E_0$ for $\Gamma = \Gamma_{\perp} = 5\,\mathrm{meV}$ and different values of
  detuning $\omega_{\mathrm{d}} = 1\,\mathrm{meV}$, 
  $5\,\mathrm{meV}$,  $10\,\mathrm{meV}$. } \label{KerrRWA_Gamma}
\end{figure}
\begin{figure}[!tb]
  \includegraphics[width=8.5cm,angle=0]{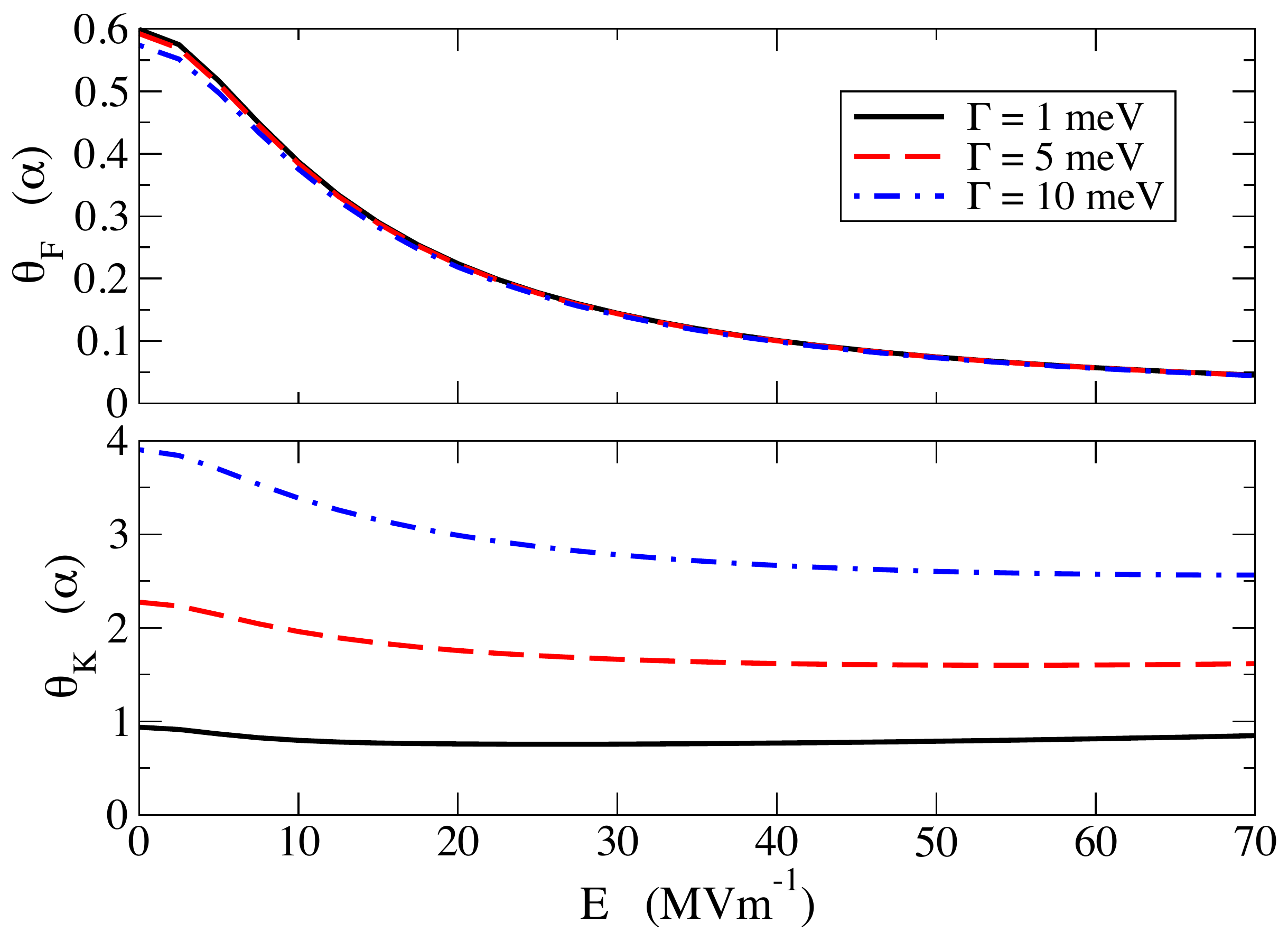}
\caption{(Color online) Faraday $\theta_F$ and Kerr $\theta_K$ angles (in units of
  $\alpha$ radians) versus electric 
  field $E_0$ for detuning $\omega_{\mathrm{d}} = 5\,\mathrm{meV}$ and
  different values of 
  $\Gamma = \Gamma_{\perp} = 1, 5, 10\,\mathrm{meV}$. } \label{FaraKerrRWA_Gamma}
\end{figure}

Figs.~\ref{FaraRWA_Gamma}-\ref{KerrRWA_Gamma} show the calculated
Faraday and Kerr rotations from the numerical solutions of
Eqs.~(\ref{TxG})-(\ref{TyG}). The decreasing trend of the Faraday
angle and the non-monotonic behavior of the Kerr angle with increasing 
electric field remain similar to the coherent case.  Although the reactive components of $\mathcal{J}_{x}$ and $\mathcal{J}_y$ are reduced by
relaxation (c.f. Fig.~\ref{JRWA_detun} and Fig.~\ref{JrtRWA_G_detun}), we find that, interestingly, the Faraday and Kerr effects 
are enhanced in the presence of relaxation and dephasing compared to
the coherent case (Figs.~\ref{FaraRWA_detun}-\ref{KerrRWA_detun}). 
In particular, $\theta_K$ is enhanced by an order of magnitude 
and is more sensitive than $\theta_F$ to increasing
values of $\Gamma$ and $\Gamma_{\perp}$, as depicted in 
Fig.~\ref{FaraKerrRWA_Gamma}. To
a lesser extent, this behavior is also observed when
$\theta_K$ is varied as a function of detuning $\omega_d$. 

To exhibit this curious enhancement and
sensitivity of $\theta_{K}$ over $\theta_{F}$, we consider the
weak-field regime when nonlinear effects are absent and evaluate the
Faraday and Kerr angles up to leading order in $\alpha = e^2/\hbar c$. From 
Eqs.~(\ref{TxLin})-(\ref{TyLin}) we have $T_x \simeq 1
-2\pi\sigma_{xx}/c$, $T_y \simeq
-2\pi\sigma_{yx}/c$, and $R_x \simeq -2\pi\sigma_{xx}/c$, $R_y \simeq
-2\pi\sigma_{yx}/c$, where $\sigma_{xx}, \sigma_{yx}$ are the
longitudinal and Hall conductivities corresponding to
Eqs.~(\ref{JxG})-(\ref{JyG}). For the transmitted light, $\mathrm{Re}\left\{T_{\pm}\right\}
\approx 1+\mathcal{O}(\alpha^1)$ and $\mathrm{Im}\left\{T_{\pm}\right\}
\approx \mathcal{O}(\alpha^1)$ imply that $\theta_F \approx
\tan^{-1}\left[\mathcal{O}(\alpha^1)\right]$. For the reflected light,
however, we have both $\mathrm{Re}\left\{R_{\pm}\right\}, \mathrm{Im}\left\{R_{\pm}\right\}
\approx \mathcal{O}(\alpha^1)$ yielding $\theta_K \approx
\tan^{-1}\left[\mathcal{O}(\alpha^0)\right]$. The Kerr
angle is therefore larger than the Faraday angle by about an order of
magnitude, resulting in the larger enhancement and increased
sensitivity with changes in relaxation rate and detuning.  


\section{Conclusion}

In summary, we have developed a theory for the magneto-optical
effects in topological insulators under intense laser fields in the small
detuning regime $\delta \ll \Delta$. We calculated the nonlinear
longitudinal and Hall currents in response to linearly polarized
light and obtained the Faraday and Kerr rotations as a function of
the incident electric field. Surprisingly, damping effects due to relaxation
and dephasing are found to enhance the resulting Faraday and Kerr
rotations. In particular, the Kerr effect exhibits a larger
enhancement and higher sensitivity to changes in detuning and damping
rate than the Faraday effect. As limiting cases, we examined the current
responses in the weak field and strong field regimes. In the weak
field regime, the currents we obtained under rotating-wave approximation
account exactly for the resonant contribution in linear 
response theory. In the strong field regime, no such correspondence
can be found. The currents saturate for negligible damping but 
decrease with electric field when damping is taken into
account. The leading-order nonlinear 
dependence of the Faraday and Kerr rotations on the incident field
implies that optical Stark effect can be probed using
magneto-optical spectroscopy. Although we focused on the case
of a topological insulator surface, our results also carry over
directly to 2D Chern insulators, whose low-energy theory is described similarly by the 2D
Dirac model.

\section{Acknowledgement}

The author acknowledges useful discussions with R. Binder.
This work is supported by a startup fund from the University of
Alabama. 

\newpage
\section{Appendix} \label{sec:append}

\subsection{Spin Representation} \label{sec:spin}

The expressions for the spin vectors defined in Sec.~\ref{sec:model} are 
\begin{eqnarray}
\hat{\boldsymbol{\alpha}}_{k} &=& 
\sin\theta_k \hat{\bm{k}}+\cos\theta_k\hat{\bm{z}}, \label{basis1} \\
\hat{\boldsymbol{\beta}}_{k} &=& \hat{\bm{z}}\times \hat{\bm{k}}, \nonumber \\
\hat{\boldsymbol{\gamma}}_{k} &=& -\cos\theta_k \hat{\bm{k}}+
\sin\theta_k\hat{\bm{z}}. \nonumber
\end{eqnarray}

This gives the following longitudinal and transverse spin vectors
defined in Sec.~\ref{sec:dressed}
\begin{eqnarray}
\hat{\boldsymbol{\alpha}}_{k}^{\mathrm{L}} &=& 
\sin\theta_k \hat{\bm{k}}+\cos\theta_k\hat{\bm{z}}, \label{basis2} \\
\hat{\boldsymbol{\alpha}}_{k}^{\mathrm{T}} &=& \hat{\bm{z}}\times \hat{\bm{k}}+i \cos\theta_k \hat{\bm{k}}-i\sin\theta_k\hat{\bm{z}},
\nonumber
\end{eqnarray}
where $\hat{\bm{k}}= \cos\phi_k\hat{\bm{x}}+
  \sin\phi_k\hat{\bm{y}}$ is the unit momentum vector with an 
  azimuthal angle $\phi_k$. 


\end{document}